\documentclass[journal]{IEEEtran}
\newcommand{\RNum}[1]{\uppercase\expandafter{\romannumeral #1\relax}}
\usepackage{mathrsfs}
\usepackage{amssymb}
\usepackage[mathcal]{euscript}
\usepackage{diagbox}
\usepackage{cite}
\usepackage[T1]{fontenc}
\usepackage{graphicx}
\usepackage{makecell}
\usepackage{psfrag}
\usepackage{url}
\usepackage{stfloats}
\usepackage{amsmath}
\usepackage{array}
\usepackage{float}
\usepackage{multirow} 
\usepackage{amsthm}
\usepackage{color}
\usepackage{multicol}
\usepackage{stfloats}
\usepackage{enumerate}
\usepackage{subfigure}
\usepackage{booktabs}  

\allowdisplaybreaks[4]
\usepackage[ruled]{algorithm2e}
\usepackage{algorithmic} 
\ifCLASSINFOpdf
\else
\fi
\begin{document}
%
\title{Optimizing Generative AI Networking: A Dual Perspective with Multi-Agent Systems and Mixture of Experts}



%
\author{Ruichen Zhang,  Hongyang Du, Dusit Niyato,~\IEEEmembership{Fellow,~IEEE}, Jiawen Kang, \\Zehui Xiong,    Ping Zhang,~\IEEEmembership{Fellow,~IEEE}, and Dong In Kim,~\IEEEmembership{Fellow,~IEEE}

\thanks{R. Zhang, H. Du, and D. Niyato are with the College of Computing and Data Science, Nanyang Technological University, Singapore (e-mail: ruichen.zhang@ntu.edu.sg, hongyang001@e.ntu.edu.sg, dniyato@ntu.edu.sg).}

\thanks{J. Kang is with the School of Automation, Guangdong University of Technology, China (e-mail: kavinkang@gdut.edu.cn).}

\thanks{Z. Xiong is with the Pillar of Information Systems
Technology and Design, Singapore University of Technology and Design, Singapore (e-mail: zehui\_xiong@sutd.edu.sg).}

\thanks{P. Zhang is with the State Key Laboratory of
Networking and Switching Technology, Beijing University of Posts and Telecommunications, China (e-mail: pzhang@bupt.edu.cn).}

\thanks{D. I. Kim is with the Department of Electrical and Computer Engineering, Sungkyunkwan University, Suwon 16419, South Korea (email:dikim@skku.ac.kr).}

}
\maketitle

\begin{abstract}
In the continued development of next-generation networking and artificial intelligence content generation (AIGC) services, the integration of multi-agent systems (MAS) and the mixture of experts (MoE) frameworks is becoming increasingly important. Motivated by this, this article studies the contrasting and converging of MAS and MoE in AIGC-enabled networking. First, we discuss the architectural designs, operational procedures, and inherent advantages of using MAS and MoE in generative AI to explore its functionality and applications fully. Next, we review the applications of MAS and MoE frameworks in content generation and resource allocation, emphasizing their impact on networking operations. Subsequently, we propose a novel multi-agent-enabled MoE-proximal policy optimization (MoE-PPO) framework for 3D object generation and data transfer scenarios. The framework uses MAS for dynamic task coordination of each network service provider agent and MoE for expert-driven execution of respective tasks, thereby improving overall system efficiency and adaptability. The simulation results demonstrate the effectiveness of our proposed framework and significantly improve the performance indicators under different network conditions. Finally, we outline potential future research directions.

\end{abstract}

\section{Introduction}
With the arrival of B5G (B5G) and 6G, a range of emerging services, including virtual reality (VR) and augmented reality (AR), will transform user experience through unprecedented data rate requirements and enhanced interaction quality. These complex applications require powerful tools to efficiently generate, manipulate, and deliver multimedia content in real-time, moving toward the artificial intelligence-generated content (AIGC) era. At the core of AIGC is generative AI (GenAI), which is able to synthesize new data instances that reflect the complex patterns of its training datasets \cite{10529221}. This includes the dynamic creation of text, image and video content through deep learning models that simulate the statistical characteristics of input data. However, the deployment of GenAI systems is not without challenges. For instance, some famous text-to-image models, such as DALL-E and GLIDE, require extensive computational resources, including 12 billion and 3.5 billion parameters, respectively\footnote{https://www.techtarget.com/searchenterpriseai/definition/Dall-E}. Such a large number of parameters results in high computational complexity and reduced adaptability, making it difficult for these systems to adapt to new or changing data scenarios quickly. The substantial resource requirements and lack of flexibility of these models highlight the need for more efficient and adaptable GenAI solutions.

To address the inherent high complexity of GenAI models, the mixture of experts (MoE) method has been proposed as a promising solution. Specifically, by building a network around dynamic mechanisms, MoE efficiently distributes computational tasks among specialized sub-models, or "experts", each fine-tuned for specific operations \cite{xue2024raphael}. This design not only effectively reduces the number of parameters, but also increases the processing speed. For example, Google’s Switch Transformers model employs 1.6 trillion parameters, but thanks to its MoE architecture, it can handle tasks efficiently using only necessary expert parts of the network at any given time \cite{fedus2022switch}. This allows for more efficient use of computing resources and faster processing times, which is critical for the high throughput required by next-generation networks.

On the other hand, to alleviate the low adaptability of GenAI models, a multi-agent system (MAS) has been proposed as another potential solution. Specifically, in MAS, multiple autonomous agents collaborate to solve complex problems dynamically \cite{yuan2024mora}. Each agent in the system operates independently but coordinates with other agents to adjust its policies based on real-time data. MAS is capable of meeting real-time decision-making environments and is particularly useful in scenarios such as network security and traffic management, where conditions are constantly changing. For example, DeepMind AlphaStar develops AI-driven gaming environments. Here, multiple agents, each of which controls a character or game element independently, analyze a game environment and make strategic decisions. These decisions are synchronized and learned through ongoing interactions, enhancing the complexity of the game and the realism of the interactions\footnote{https://deepmind.google/discover/blog/alphastar-grandmaster-level-in-starcraft-ii-using-multi-agent-reinforcement-learning/}.
 
Inspired by these considerations, our work proposes a hybrid framework that integrates the advantages of MoE and MAS to overcome the limitations of stand-alone GenAI systems. By combining the efficient computational management of MoE with the dynamic adaptability of MAS, this framework aims to enhance the generative capabilities and operational flexibility of GenAI applications. In this work, our main contributions are summarized as follows:
\begin{itemize}

\item We explore the architectural design, operational workflows, and advantages of MAS and MoE in the context of GenAI, studying their functional integration.
\item We review the main applications of MAS and MoE for GenAI, focusing on content generation and resource allocation and emphasizing their impact on next-generation networking.

\item We propose a novel multi-agent-enabled MoE framework for a novel 3D object generation and data transfer scenarios. The framework uses MAS for dynamic task coordination of each network service provider agent and MoE for expert-driven execution of respective tasks, thereby improving overall system efficiency and adaptability.

\end{itemize}

\section{Multi-agent and Mixture of Experts for GenAI}

\begin{figure*}[!t]
\centering
\includegraphics[width=0.85\textwidth]{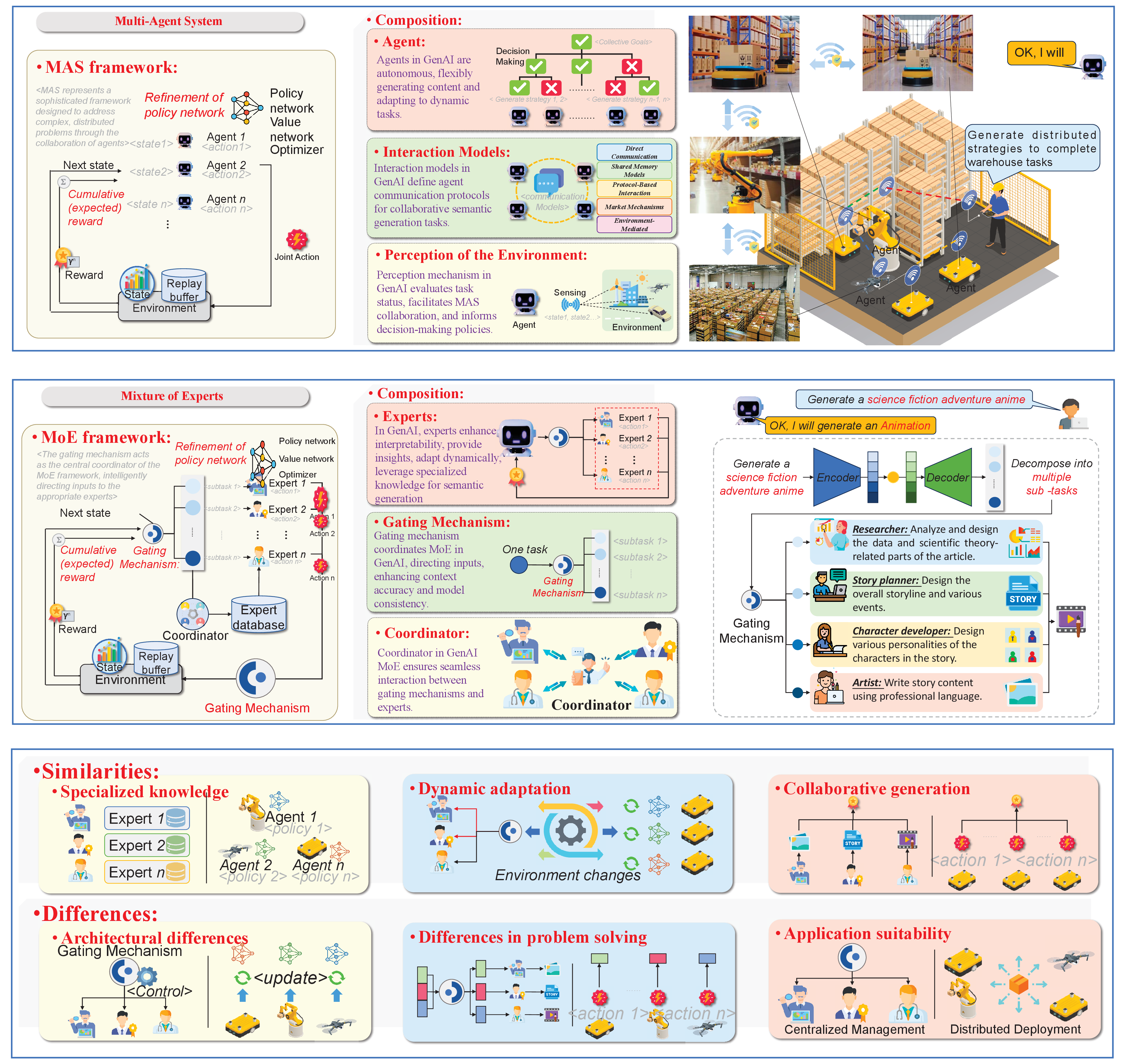}
\caption{Structure and function of MoE and MAS for GenAI. MoE framework is described as a dynamic network where a central gating mechanism delegates tasks to specialized experts, focusing on feedback for performance optimization and benefits such as precision and customization. MAS is described as a collaborative environment where agents use shared protocols for interaction, decision-making, and context awareness.}
\label{fig_PSNC_1}
\end{figure*}

In this section, we explore the structures, working processes, advantages, and contrasts of multi-agent systems (MAS) and MoE for GenAI, where a diagram is shown in Fig.~\ref{fig_PSNC_1}.

\subsection{Overview of GenAI}

GenAI leverages generative learning techniques that enable AI models to not only learn from data but also to generate new instances that accurately reflect complex patterns and structures in the training data set. This generative capability is useful for applications requiring novel content creation, predictive modeling, and data augmentation. Key GenAI models include variational autoencoder (VAE), generative adversarial network (GAN), and diffusion-based model (DBM), each of which offers unique generative mechanisms \cite{10529221}. In particular,  VAE is capable of structured output and handling of missing data, which is good at handling high-dimensional data tasks. GAN uses generative and discriminative models to improve iteratively through competition, thereby increasing the fidelity of generated instances. DBM is generated by simulating a denoising process that converts random noise into coherent structures over time.

Despite these unique generation capabilities, GenAI models face several limitations in their current implementations, which are mostly based on the centralized architecture:
\begin{itemize}

    \item \textbf{Unmanageable Model Complexity:} GenAI is highly complex because of its extensive parameter set and the need for significant training datasets. This complexity requires considerable computing resources and complicates the model training and fine-tuning processes. Substantial resource requirements can create barriers to deployment, especially in resource-constrained environments\footnote{https://itrexgroup.com/blog/calculate-the-cost-of-generative-ai/}.

    \item \textbf{Low Adaptability:} GenAI is less adaptable due to its reliance on predefined datasets and algorithms\cite{wang2024toward}. This dependency, coupled with the large scale of the models, limits their flexibility in responding to new or changing data patterns and downstream tasks.

   \item \textbf{Performance Bottleneck:} GenAI has to process and integrate datasets or model parameters onto a central server. This centralized approach can create severe bottlenecks, especially regarding data transfer speed and processing time\cite{10529221}. 
\end{itemize}

\subsection{Multi-Agent System for GenAI}
MAS represents a sophisticated framework designed to address complex, distributed problems through the collaboration of agents \cite{yuan2024mora}. Considering semantic extraction and generation tasks of GenAI, MAS can leverage multiple autonomous agents that run independently and then collaborate in a shared environment to generate policies to handle semantic generation tasks in complex networks.

\subsubsection{Principles and Composition}
The core principle of MAS is agent autonomy, which enables the system to respond flexibly to dynamic generation tasks in the decentralized structure. Agents operate independently, making decisions and taking actions based on their unique perceptions, goals, and information acquired from the environment. Thanks to this decentralized mechanism, agent independence is crucial in semantic generation scenarios. For example, one agent might specialize in decoding and extracting text from visual input such as images or videos, while another agent works on creating descriptive narratives or generating informative content based on the extracted information. This further facilitates more adaptive and personalized content generation that incorporates user preferences. Furthermore, within MAS, agents are coordinated through well-defined interaction models and protocols, ensuring efficient collaboration and resource allocation to achieve complex semantic generation tasks.

For MAS, it consists of several key components, i.e., agents, interaction models, and perception of environments. Each component contributes to the functionality and efficiency of semantic generation, i.e.,
\begin{itemize}
    \item \textbf{Agents:} In GenAI for semantic generation, agents are self-directed and have generative capabilities, enabling them to generate new content or solutions independently. Their autonomy enables MAS to respond flexibly and adaptively to dynamically generated tasks and environmental conditions. For example, in \cite{zou2003using}, agents adopted their respective Semantic Web technologies to customize the best travel package by generating and reasoning about the customer's semantic data to meet their specific requirements, such as budget, preferences, and schedule. Through agents, MAS can respond flexibly and adaptively to dynamically generated tasks and environmental conditions.

    \item \textbf{Interaction Models:}  In GenAI for semantic generation, interaction models are the basis, which define the protocols for how agents communicate and collaborate on generation tasks.  For example, in \cite{zhang2023optimization},  the interaction mechanism facilitated collaboration between servers to jointly transmit image data to groups of users by using semantic communication techniques. This enabled the server to specifically transmit semantic information that accurately captures the core meaning of the image, effectively generating a semantically rich representation of the raw image data. Through interaction models, MAS can negotiate the partitioning of generative tasks and coordinate actions to efficiently achieve complex generative goals.

    \item \textbf{Perception of the Environment:} In GenAI for semantic generation, the perception mechanism is essential because of the need to identify the current status and requirements of the generation task, evaluate the progress of content creation, and understand the behavior of other agents. For example, in \cite{yuan2024mora}, the perception mechanism facilitated multi-agent collaboration by observing an environment workflow from text interpretation to complex video generation, ensuring a coherent and high-quality semantic generation process and implementing image, image-to-video generation. Through the perception mechanism, MAS can effectively integrate environmental data and internal state information to generate corresponding decision-making policies.
\end{itemize}

\subsubsection{Advantages of MAS for GenAI}

Integrating MAS into GenAI applications introduces substantial advantages that directly address the limitations of centralized GenAI approaches. Firstly, MAS enhances operational efficiency by parallel processing tasks by distributing different components of the generative process among specialized agents\cite{zhang2023optimization}. For instance, while one agent might generate textual content in semantic content generation, another could simultaneously work on visual synthesis, significantly accelerating the generation process. Secondly, the diverse strategies and methods employed by individual agents in MAS lead to the creation of contents that are rich in variety and tailored to a broader spectrum of user preferences. This diversity, coupled with MAS's inherent robustness wherein the failure of a single agent does not compromise the entire system's functionality ensures reliable service delivery for critical and large-scale generative tasks. Moreover, MAS's adaptability allows for dynamic adjustments in response to evolving environmental conditions or user feedback, ensuring the generated content remains highly relevant.

\subsection{Mixture of Experts for GenAI}
MoE is an architecture that integrates a set of specialized neural network components, i.e., experts, to handle the given tasks, with each expert fine-tuned to handle specific types of sub-tasks or sub-datasets \cite{xue2024raphael}.

\subsubsection{Principles and Compositions}
The core principle of MoE framework is to operate according to what each expert model is good at \cite{zhang2024interactive}. Specifically, each expert, which may include generative AI models, is trained to focus on a specific subset of the problem space, leveraging their strengths to provide specialized solutions. The overall strategy of MoE is to dynamically select and combine the insights of these experts based on the task at hand, guided by a gating mechanism that determines the relevance and weight of each expert's contribution to the final outcome. For example, a semantic generation task of GenAI may require one expert to focus on extracting topics and sentiments from large amounts of text data while another expert is responsible for producing corresponding content with these extracted topics, ensuring high relevance and engagement. Furthermore, the gating mechanism evaluates specific requirements of the input content to determine suitable experts and strategies. In this case, the gating network will route the input related to the topics and sentiments to the first expert while the input related to content generation is sent to the second expert.

\begin{table*}[!t]
\centering
\caption{Summary of similarity and differences between MAS and MoE for GenAI.} 
\includegraphics[width=0.85\textwidth]{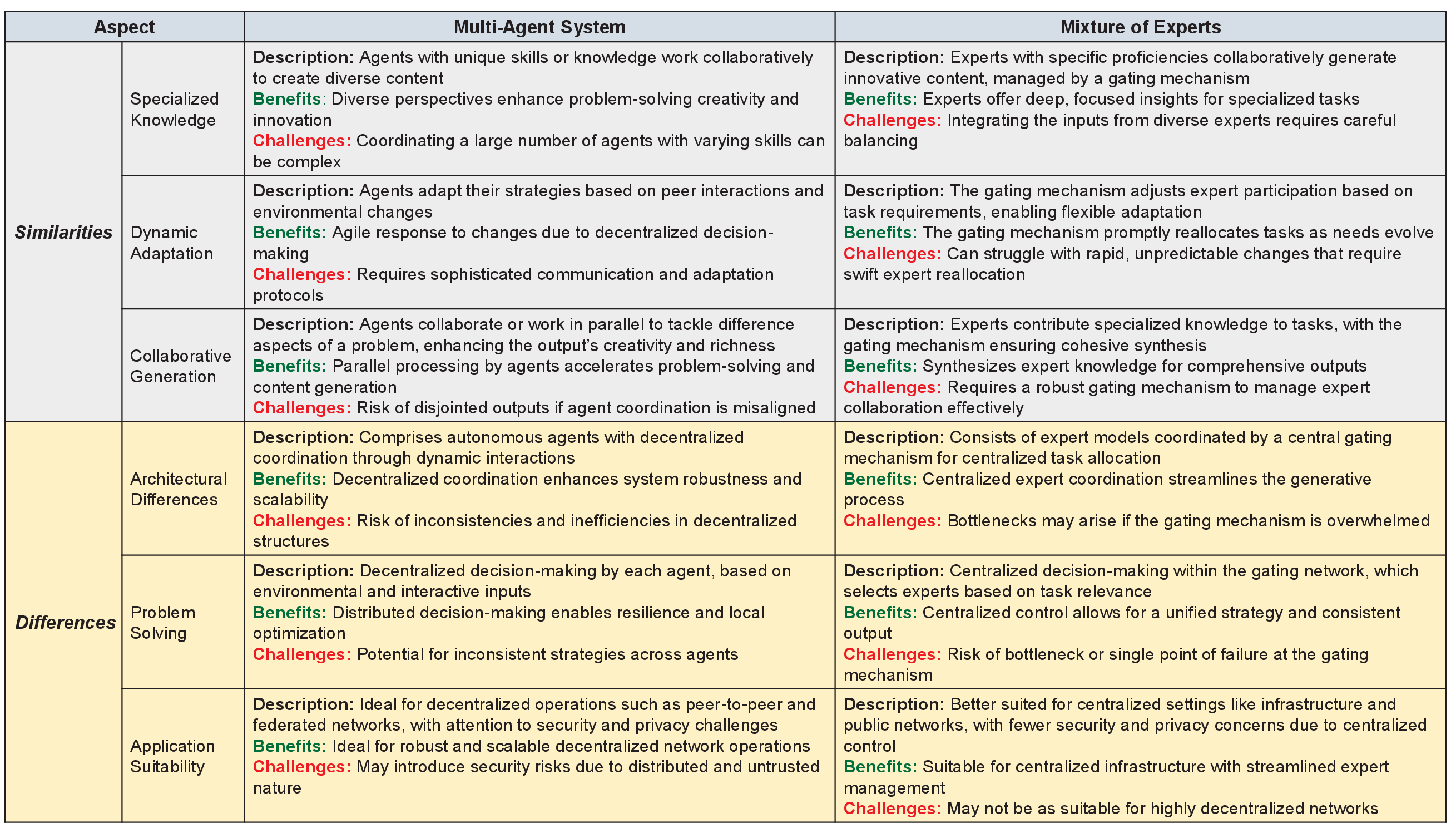}
\label{MAS_TABLE}
\end{table*}

For MoE, it consists of several key components, i.e., experts, gating mechanism, and coordinator.  Each component contributes to the functionality and efficiency of GenAI, i.e.,
\begin{itemize}
      \item \textbf{Experts:} In GenAI for semantic generation, experts are models tailored to specific parts of the problem space, which includes semantic generative tasks such as semantic extraction and generation. For example, in \cite{pavlitskaya2020using}, by dividing the input data into semantic subsets relevant to different driving scenarios, each expert focuses on their respective areas, enhancing the interpretability of autonomous driving neural networks by providing insights into the decision-making process through their agreement or disagreement. Through experts, MoE can dynamically adapt to diverse semantic contexts in GenAI, enabling the system to accurately model and generate content across a wide range of scenarios by effectively leveraging different experts' specialized knowledge and skills.

      \item \textbf{Gating Mechanism:} In GenAI for semantic generation,  the gating mechanism acts as the central coordinator of MoE framework, intelligently directing inputs to the appropriate experts, including when the generative capabilities of the GenAI component are required. It evaluates the specifics of each task, including its context and requirements, to determine which experts are best suited to contribute to the solution. For example, in \cite{pavlitskaya2020using}, gating mechanisms can intelligently coordinate expert contributions to semantic generation, making granular decisions about which expert to use based on the context and requirements of each driving scenario. This process not only ensures that context-accurate semantic interpretations are generated for autonomous driving, but also enhances model consistency by highlighting areas of agreement or disagreement among experts.

      \item \textbf{Coordinator:} In GenAI for semantic generation, although the coordinator is not a standard component of all MoE implementations, it plays a vital role when present. It coordinates interactions between gating mechanisms and experts, ensuring seamless communication and processes. For example, in \cite{10191738},  the coordinator can effectively manage the interactions between gating mechanisms and experts, ensuring that the right expert is selected for processing based on the specific semantics of the input text. This coordination is crucial for leveraging the nuanced understanding of entities and events within sentences, allowing for more accurate event detection and type classification by harmonizing the adaptive semantic encoding process.
\end{itemize}

\subsubsection{Advantages of MoE for GenAI}
MoE framework markedly enhances GenAI tasks by integrating specialized expertise for semantic generation, streamlining the creation process while ensuring high-quality outputs. Through its dynamic gating mechanism, MoE adeptly assigns tasks to the most suited experts, significantly boosting efficiency and ensuring the reliability of the generated content. This strategic allocation optimizes task performance and heightens the accuracy of outputs, as it leverages the collective strengths and specialized knowledge of various experts\cite{wang2024toward}. Moreover, MoE's centralized control mechanisms enhance security, meticulously regulating data access among experts to protect sensitive information and adhere to privacy standards. Together, these features position MoE as a potent architecture for advancing the capabilities and applications of GenAI, combining efficiency, reliability, and security in one cohesive framework.

\subsection{Similarities and Differences}
We summarize the similarities and differences between MAS and MoE for GenAI, clarifying how each framework facilitates content generation. For similarities, both MAS and MoE leverage expertise, with MAS employing autonomous agents and MoE utilizing designated experts to provide diverse and innovative solutions to complex problems. This specialization fosters collaborative production of contents in GenAI, with each unit contributing unique insights to create rich, multifaceted outputs. Additionally, their dynamic adaptability ensures that both frameworks can respond to changing requirements, optimizing performance through continuous adjustments.

Despite these similarities, MAS and MoE differ significantly in structure and operational paradigms. MAS is characterized by decentralization, with agents navigating independently and responding to environmental cues, promoting a distributed approach to problem solving. Instead, MoE operates under a centralized coordination mechanism, with a gating system that allocates tasks precisely to the most appropriate experts. This centralized decision-making process is particularly beneficial in structured environments, providing simplified management and potentially enhanced security. For clarity, similarities and differences between MAS and MoE for GenAI are summarized as Table~\ref{MAS_TABLE}.

\section{Applications of MAS and MoE for GenAI in networking}

In this section, we explore applications of MAS and MoE for GenAI in networking, including content generation and resource allocation, where the corresponding related studies are summarized in Table~\ref{works_TABLE}.
\begin{table*}[!t]
\centering
\caption{Summary of related works of MAS and MoE for GenAI.} 
\includegraphics[width=0.85\textwidth]{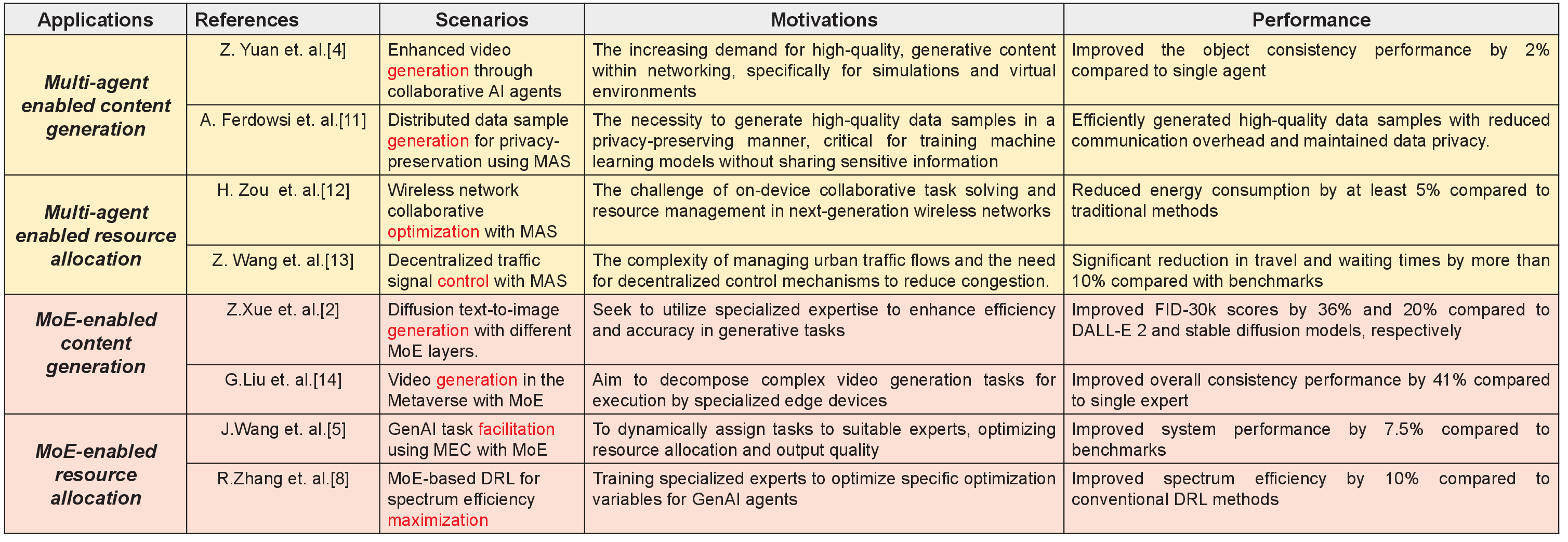}
\label{works_TABLE}
\end{table*}


\subsection{Applications of MAS for GenAI}

\subsubsection{Multi-agent enabled GenAI content generation}
Content generation by GenAI services requires the creation of user content, business information, and scenarios that are critical for testing and simulating. Here, MAS-enabled GenAI is able to generate these elements in a distributed manner, thereby increasing generation efficiency and effectiveness. For example, in \cite{yuan2024mora}, the authors introduce Mora, a framework to improve collaboration among multiple AI agents for video generation tasks. It provides a system for configuring components and task pipelines, enabling the integration of multiple AI agents specialized in text-to-image, image-to-image, image-to-video, and video-to-video transformations. Moreover, in \cite{10265048}, the authors introduce the brainstorming GAN (BGAN) architecture for generating high-quality data samples, a multi-agent GAN solution for scenarios where data is distributed across multiple agents with a need to learn data distributions independently without sharing their datasets. BGAN enables fully distributed operation, allowing agents to exchange generated data samples instead of actual data, effectively reducing communication overhead and preserving data privacy.

\subsubsection{Multi-agent enabled GenAI resource allocation}
Resource allocation is a key aspect of GenAI services in networking. It involves strategically allocating network resources (i.e., bandwidth, computing power, and storage) among various users and applications to maximize the efficiency and satisfaction of GenAI tasks.  Here, MAS can provide dynamic resource allocation approaches that enable the policies to respond adeptly to changing network demands and conditions, optimizing the infrastructure that supports GenAI tasks. For example,  in \cite{zou2023wireless},  the authors investigate the integration of generative large language models (LLMs), edge networks, and MAS for wireless networks, focusing on collective intelligence and edge-based decision-making. They demonstrate the incorporation of multi-agent GenAI, aiming to develop on-device LLMs for multi-collaborative task solving and network objectives.  Furthermore, in \cite{9647926},  the authors propose a decentralized adaptive traffic light signal control (ATSC) framework for vehicular traffic networks, utilizing a GAN for traffic data recovery and a multi-agent DRL for signal control. Specifically, the GAN recovers traffic data for intersections in the vehicular traffic network from limited traffic statistics. The proposed multi-agent DRL algorithm allows each intersection to independently manage traffic light signals while collaboratively optimizing traffic flow across the network.   

\subsubsection{Lesson learned}
From the above applications, we learn that GenAI services can improve performance metrics and operational efficiency by harnessing MAS's adaptive and distributed learning capabilities. For example, the Mora framework illustrates the ability of MAS to facilitate rapid, innovative video and data sample generation, ensuring the creativity and security of GenAI-driven output. Additionally, the distributed nature of MAS simplifies resource allocation and significantly improves equipment intelligence and transportation system efficiency. Despite these advances, integrating MAS with GenAI services is not without challenges. The independence of agents in MAS sometimes leads to a fragmentation of task focus, which is particularly evident in the fragmented execution of GenAI tasks. To solve this problem, centralized coordination of MoE can provide more unified goals. 


\subsection{Applications of MoE for GenAI}

\subsubsection{MoE-enabled content generation}
MoE is able to enhance the efficiency and accuracy of GenAI content generation processes by utilizing specialized experts for different aspects of a task.  For example, in \cite{xue2024raphael}, the authors introduce RAPHAEL, a text-to-image generation framework incorporating a diffusion model with space-MoE and time-MoE layers. Specifically,  RAPHAEL first uses self-attention and cross-attention to integrate textual and visual input. The space-MoE layer then maps textual concepts to corresponding image segments, while the time-MoE layer modulates the temporal aspects of image formation. Ultimately, a text-gating network aggregates these operations, ensuring the output is a unified visual representation of the initial text input. Moreover, in \cite{liu2024fusion}, the authors propose a framework integrating MoE with GenAI for video generation in the metaverse. Specifically, the framework starts with LLM, splitting the video generation task into smaller, manageable tasks. Subsequently, specialized edge devices (each an expert in a different aspect of temporal or spatial video generation) perform their assigned subtasks. Finally, these parallel tasks are managed by a gating network to produce the final videos.

\begin{figure*}[!t]
\centering
\includegraphics[width=0.85\textwidth]{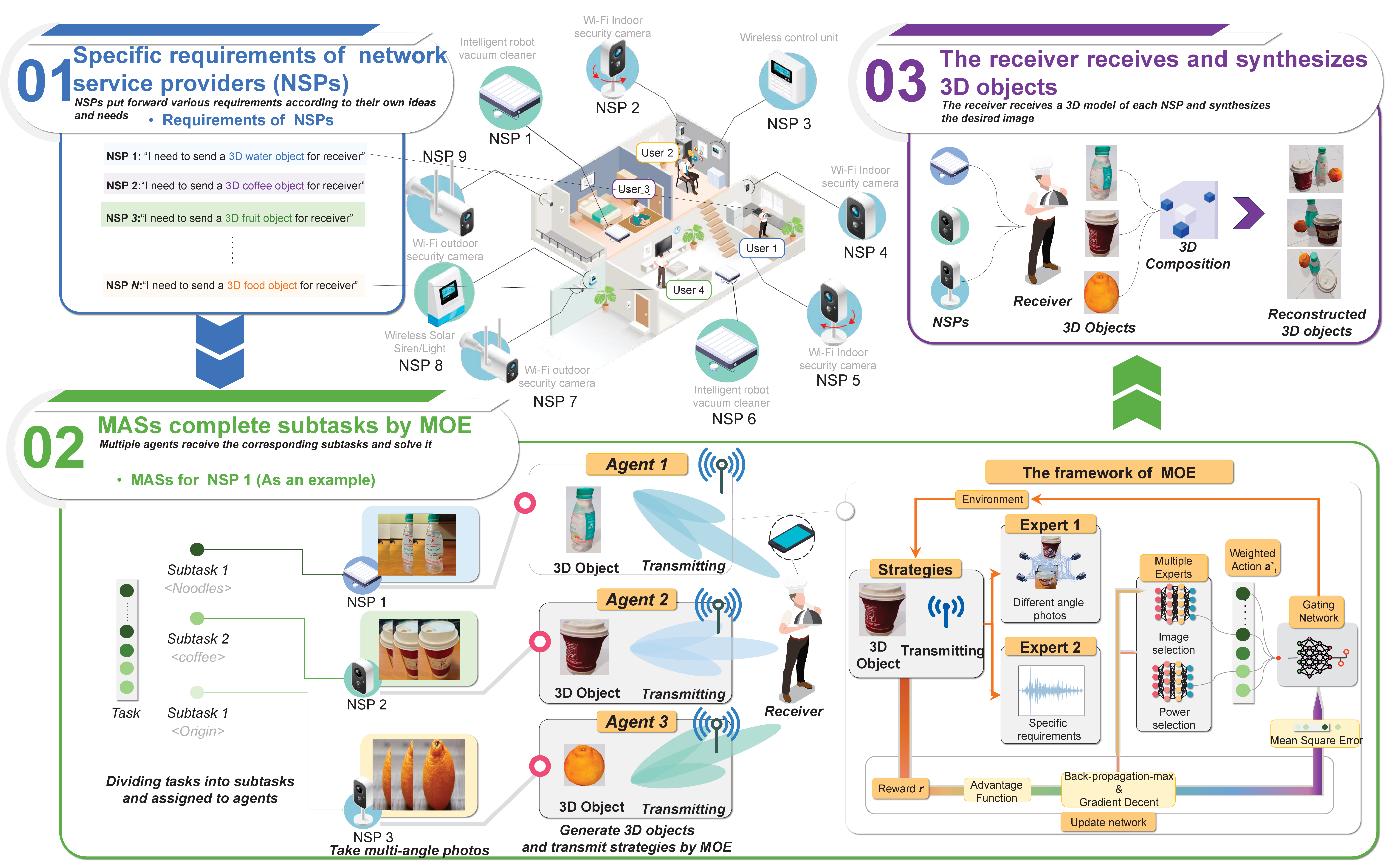}
\caption{The Structure of system model and proposed solution framework. Multiple NSPs (each NSP responsible for a different type of food) collaborate to collect, process, and tramsit 3D objects of food. The left part of the figure shows the process from collecting various 2D images from different angles for each NSP to generate these images into a 3D object. The right part introduces MoE framework, in which multiple experts, guided by a gating network, optimize specific subtasks, such as image selection and power allocation, to improve efficiency and reduce overall transmission costs.}
\label{fig_PSNC_2}
\end{figure*}

\subsubsection{MoE-enabled resource allocation}
MoE is able to optimize GenAI resource allocation by dynamically assigning tasks to the most suitable experts, enhancing both efficiency and output quality.  For example,  in \cite{wang2024toward}, the authors propose a framework that leveraged mobile edge networks to facilitate GenAI tasks on edge devices. When prompted by the edge device, the process begins by splitting the build task into subtasks. The framework then evaluates the required computing resources and coordinates the assignment of these subtasks to different edge experts, considering wireless channel conditions and expert availability. A DRL-based algorithm is proposed to optimize this allocation to select edge experts based on their execution capabilities.  Moreover, in \cite{zhang2024interactive}, the authors introduce an MoE-based DRL method to solve the spectrum efficiency maximization problem generated by  GenAI agents. This method first trains different experts, and then each expert is specially designed to optimize specific optimization variables such as beamforming vectors and common rate. Next, these optimized variables are aggregating through a gating network to coordinate the joint optimization process.
\subsubsection{Lesson learned}
From the above applications, we learn that MoE enhances GenAI for content generation and resource allocation in networking environments. For instance, within the MS-COCO 256x256 dataset, the RAPHAEL framework, employing centralized coordination via MoE, improves the FID-30k scores by 36\% and 20\% compared to DALL-E 2 and stable diffusion models, respectively.  Furthermore, thanks to its efficient gating mechanism, MoE-DRL method approach enhances system performance by approximately 10\% compared to conventional DRL methods. However, the centralized nature of MoE might introduce latency issues, particularly when coordinating a large number of expert inputs across complex tasks, potentially slowing down the response time. By decentralizing decision-making and leveraging local processing, MAS can mitigate some of the latency issues inherent in centralized systems.


\section{Case study: 3D objects generation and transmission based on MAS and MoE  }

\subsection{Motivation and System Model}
Dynamic content creation and delivery are important functions of a number of applications. MAS and MoE frameworks play a crucial role in enhancing these processes across various scene transfers, including the domain of 3D object creation and transmission. \cite{qian2023magic123}. Specifically, traditional methods of transmitting multiple 2D images are becoming insufficient. Not only do these methods consume excessive bandwidth, but also they fail to deliver the immersive experience that users expect. For example, transferring three to five images can consume up to 20MB of data, while generating a single 3D object representation through the AIGC service only requires about 7MB. Using 3D objects not only reduces bandwidth usage but also significantly enhances the viewing experience by providing more realistic and interactive visualizations. Here, MAS coordinates the work of individual agents to efficiently collect, process, and transmit data. Simultaneously, MoE leverages domain-specific expertise to optimize the quality and fidelity of generated 3D objects. Together, these frameworks ensure that 3D representations are not only bandwidth efficient but also detailed and of high quality.

Following this motivation, we consider a system model that leverages MAS integrated with MoE for 3D object generation and semantic data transfer. Fig.~\ref{fig_PSNC_2} illustrates multiple network service providers (NSPs) operating as agents at the transmitter end. These agents collect images of products (e.g., oranges, coffee, and water) from different angles under different lighting conditions to enrich visual information for generating 3D objects. In this framework, each agent is specifically responsible for the generation and transfer of different 3D objects and satisfies the constraints related to the 3D generation and transmission process. The model involves two experts at each agent. The first expert selects images that best contribute to efficient 3D generation and converts them into point cloud data. Simultaneously, the second expert is responsible for transmitting these 3D point cloud data. This transmission process is designed to minimize power consumption while maintaining a minimum data rate, ensuring that the quality of the transferred 3D object is closely related to the quality of the original model at the receiving end. Based on the system described above, our optimization problem aims to minimize the total operating costs associated with processing selected images and the power consumption costs associated with their transmission. This requires optimizing multiple parameters, including a binary decision variable for each agent to select images for 3D generation and a continuous decision variable representing each agent's transmit power. The constraints for this optimization problem include transmission power budget, image selection criteria, 3D quality contribution, and information quality of service.


\subsection{Proposed Method}
To address the described problem, we propose a multi-agent-enabled MoE-proximal policy optimization (MoE-PPO) framework. Here, each NSP is regarded as an independent agent with two experts, one for image selection and another for data transmission. This ensures arrangements for the management of every stage of the 3D object generation process, from image selection to final object transfer. Furthermore, our framework adopts centralized training and distributed execution, where each agent has its own action space, reward function, and state space. The integration of expert policies is managed through a central gating network, ensuring that decisions made by individual agents are consistent with overall system goals. This structured approach allows each agent to update its actor-network and critic-network both with all the environment state. Additionally, since all agents' states will be considered, we use max-propogation to compute the largest generalized advantage estimation to update all the networks to speed up the convergence. The action space, the state space, and the reward function are designed as follows.

\textbf{Action space:} The action space of each agent consists of two experts, one responsible for selecting images for 3D generation and one responsible for controlling the transmit power.

\textbf{State space:} The state space of each agent consists of the current action (i.e., image selection and power level), quality indicators such as SINR of the transmission model, and other environmental states that affect the decision-making, which means the state of each agent's decision-making is related to markov decision process.

\textbf{Reward function:} The reward function is to minimize the total operating cost. These costs include the computational overhead of processing the selected images and the power consumption associated with their transmission. Therefore, the reward function is inversely proportional to the cost function, increasing the reward at a lower cost. To satisfy the corresponding constraints, the reward function is combined with a penalty that is reduced to zero whenever any constraint is violated.

\begin{figure*}[!t]
\centering
\includegraphics[width=0.83\textwidth]{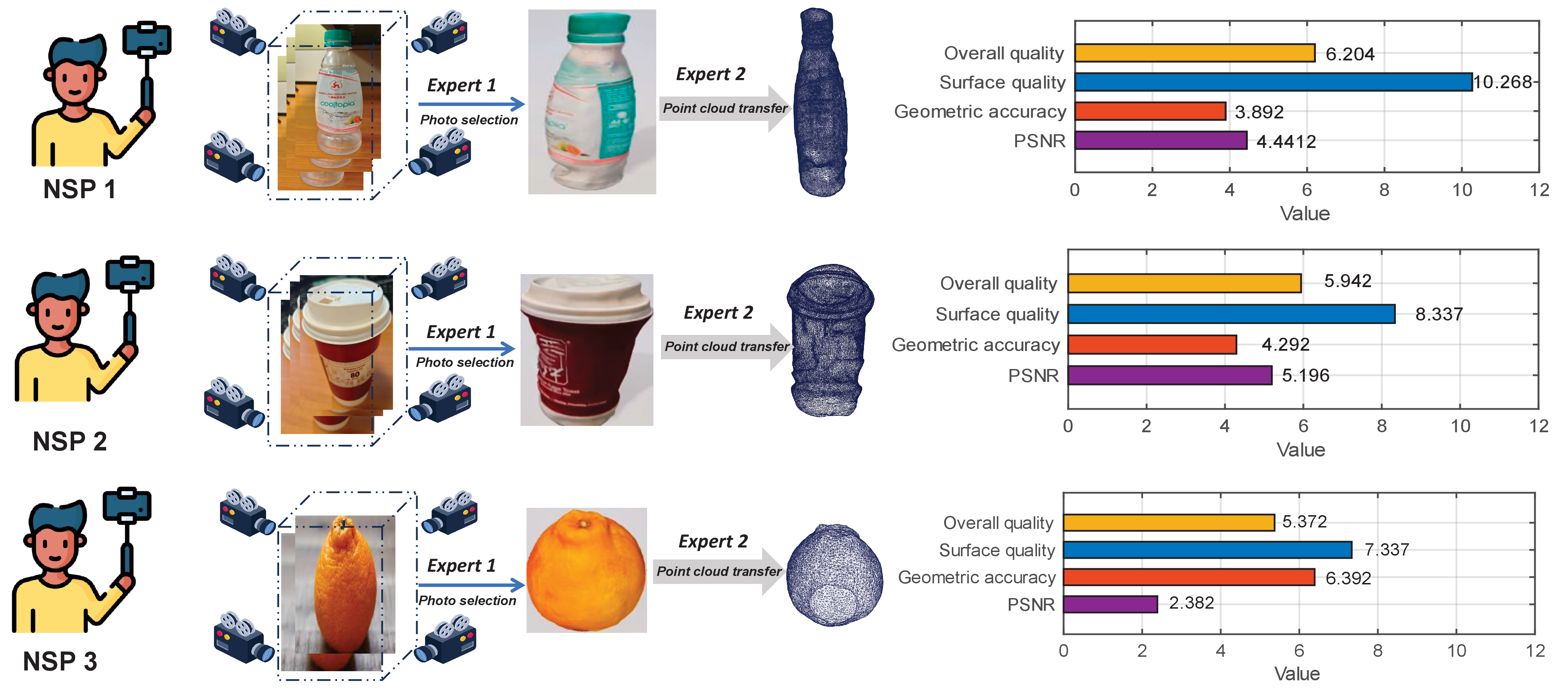}
\caption{ The efficacy of the multi-agent-enabled MoE framework for the 3D object generation and transmission system. Each NSP operates as an independent agent, with the first expert responsible for selecting the most conducive images for 3D reconstruction. These selected images are transformed into 3D objects, transferred as point clouds by a second expert specializing in data transmission. }
\label{fig_PSNC_3}
\end{figure*}

For the simulation setup, each agent is configured to select at least three images, ensuring a diverse and robust dataset for 3D objecting. These images are selected based on their significant contribution to the overall quality of the 3D generation. In terms of transmission power settings, each NSP's maximum transmit power is 20 $W$.  Our proposed multi-agent MOE-PPO framework is trained with a learning rate of \(3 \times 10^{-4}\), a clip epsilon of \(0.2\), and runs over \(15,000\) episodes across \(8\) epochs to optimize the decision-making process effectively. These settings are critical for achieving an optimal balance between computational efficiency and high-quality 3D image generation.

Fig.~\ref{fig_PSNC_3} represents the performance of the proposed framework for 3D object generation and transmission system, where each NSP is regarded as an independent agent. Each NSP utilizes the first expert to select the best images that effectively facilitate 3D reconstruction. Selected images are processed to generate 3D objects of individual items, which are transferred as point clouds using the second expert. The quality of the 3D object received at the destination is evaluated using several metrics, i.e., PSNR, geometric accuracy, surface quality, and overall quality. For example, NSP 2 focuses on coffee cups with a PSNR of 5.196, geometric accuracy of 4.292, surface quality of 8.337, and an overall quality score of 5.942. These results demonstrate the system's ability to maintain high standards of geometric accuracy and surface quality, which are crucial for generating realistic 3D objects. Furthermore, the differences in performance metrics of different NSPs highlight the importance of expert selection and image processing strategies tailored to specific types of objects to optimize the quality of the transferred 3D objects.

\begin{figure}[!t]
\centering
\includegraphics[width=0.45\textwidth]{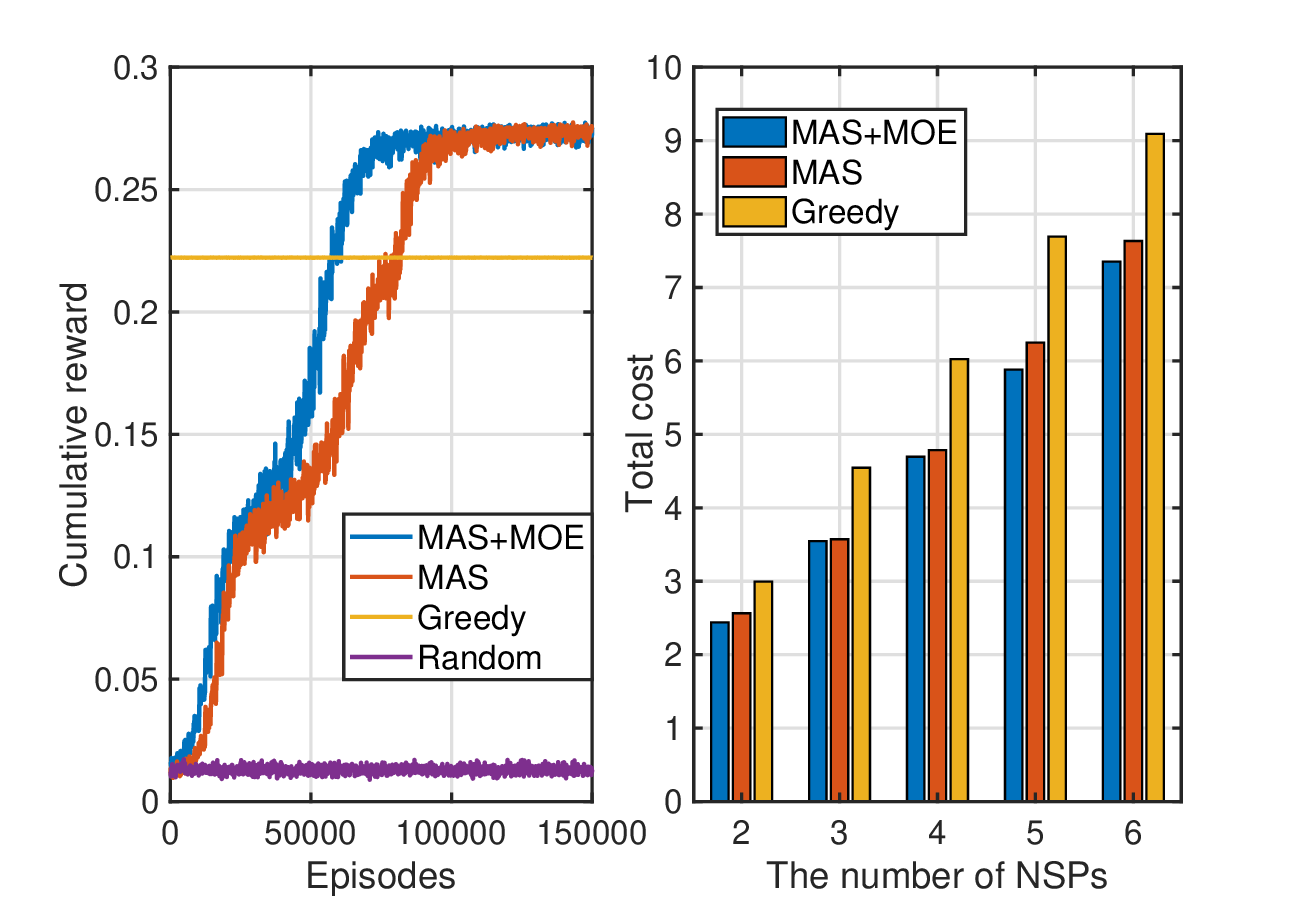}
\caption{(a) The cumulative reward versus the number of episodes; (b) The cost consumption versus the number of NSPs.}
\label{RL_MoE_1}
\end{figure}
Fig.~\ref{RL_MoE_1}(a) illustrates the convergence behavior of our proposed multi-agent-enabled MoE-PPO approach compared to various benchmarks, including multi-agent-enabled PPO, a greedy-based strategy, and a random-based strategy.  It shows that the cumulative reward of our approach gradually increases and surpasses both the greedy and random strategies.  This superior performance is attributed to PPO effectively balancing exploration and exploitation through the policy gradient method. The combination of multiple agents further enhances this by allowing a more diverse sampling of the action space, which in turn leads to more robust estimates of the policy gradient. Moreover, it also shows that although the multi-agent PPO method also converges to a similar performance to our proposed method, it requires more episodes (approximately more than 40\%). This faster convergence speed is due to MoE structure, which optimizes task distribution among specialized expert networks. Each expert in MoE configuration is trained to efficiently handle specific subtasks, thereby reducing computational overhead and accelerating the learning process for their respective responsibilities. Next, Fig.~\ref{RL_MoE_1}(b) shows the total cost of various strategies. It shows that as the number of NSPs increases, the total cost increases constantly because more NSPs process and transfer a larger volume of 3D objects, which naturally increases the overall cost. However, note that our approach consistently results in the lowest costs under all conditions. The reason may be the synergy of MAS and MoE methods, which optimize data selection and processing, as well as transfer tasks across multiple agents. Specifically, MAS framework facilitates distributed data processing, where each agent operates independently but cooperatively, reducing bottlenecks typically associated with centralized systems. At the same time, MoE architecture leverages specialized expert systems that are fine-tuned for specific subtasks in the 3D object processing pipeline, thereby minimizing redundant computation and increasing overall cost efficiency.

\section{Future Directions}
\textbf{Enhanced Multi-modal Integration:}
Future research could explore the development of multimodal MAS and MoE systems to integrate different data types (text, images, videos, etc.) better. These systems will leverage the strengths of MAS to handle dynamic wireless environments and the specialization of MoE to handle specific patterns efficiently.

\textbf{Autonomous and Optimal Adaptation Strategies:}
Future research could investigate autonomous adaptation strategies where MAS can dynamically recalibrate MoE components in response to changing environmental conditions or mission requirements. This could enhance the network system’s ability not only to react to changes but also to proactively adapt its operating strategies, thereby increasing the efficiency and effectiveness of different applications such as smart cities or adaptive security systems.

\textbf{
Semantic Communications for Complex Environments:}
Future research may explore the application of MAS and MoE frameworks in semantic communication, focusing on improving the efficiency of data interpretation and meaning transfer between devices in complex network environments such as smart cities and autonomous vehicles. This research aims to optimize transmission efficiency to ensure that transmitted data retains its intended semantic content, reduces ambiguity, and improves reliability.

\section{Conclusion}
In this article, we have introduced the structures and advantages of MAS and MoE for GenAI. Then, we have summarized some applications of MAS and MoE for GenAI in networking. In our case study, we have proposed an MAS-MoE framework to solve the 3D object generation and transmission problem, where the effectiveness of the framework has been verified through simulation results. Finally, we have outlined the potential future directions.

\bibliographystyle{IEEEtran}
\bibliography{mylib}

\end{document}